\documentclass[sigconf,authorversion]{acmart}

\usepackage{booktabs} 
\usepackage{subfig}
\usepackage{mathrsfs,amsmath}
\usepackage{array}

\usepackage{multirow}
\usepackage{amsmath}
\usepackage{xcolor}
\newcommand*{\rot}[2]{
\rotatebox[origin=l]{#1}{#2}
}
\newcommand{\teasorImgHeight}{5.6cm}

\newcommand{\HeightSimCompVis}{3.cm}
\newcommand{\rev}[1]{#1}
 
\newcommand{\volDataHeight}{4.5cm}

\newcommand{\volDataWidthCm}{5.4cm}
\newcommand{\imgDataWidth}{0.35\linewidth}

\newcommand{\tomatoHeight}{3.2cm}
\acmPrice{15.00}

\copyrightyear{2019}
\acmYear{2019}
\acmConference[SAP '19]{ACM Symposium on Applied Perception 2019}{September 19--20, 2019}{Barcelona, Spain}
\acmBooktitle{ACM Symposium on Applied Perception 2019 (SAP '19), September 19--20, 2019, Barcelona, Spain}
\acmPrice{15.00}
\acmDOI{10.1145/3343036.3343133}
\acmISBN{978-1-4503-6890-2/19/09}
\setcopyright{acmcopyright}

\setcitestyle{square}

\begin{document}

\title{Spectral Visualization Sharpening}

\author{Liang Zhou}
\affiliation{%
  \department{SCI Institute}
  \institution{University of Utah}}

\author{Rudolf Netzel}
\affiliation{%
	\department{VISUS}
	\institution{University of Stuttgart}}

\author{Daniel Weiskopf}
\affiliation{%
	\department{VISUS}
  \institution{University of Stuttgart}}

\author{Chris R. Johnson}
\affiliation{%
  \department{SCI Institute}
  \institution{University of Utah}}

\renewcommand{\shortauthors}{Zhou, Netzel, Weiskopf, and Johnson}

\begin{abstract}
In this paper, we propose a perceptually-guided visualization sharpening technique.
We analyze the spectral behavior of an established comprehensive perceptual model to arrive at our approximated model based on an adapted weighting of the bandpass images from a Gaussian pyramid.
The main benefit of this approximated model is its controllability and predictability for sharpening color-mapped visualizations.
Our method can be integrated into any visualization tool as it adopts generic image-based post-processing, and it is intuitive and easy to use as viewing distance is the only parameter.
Using highly diverse datasets, we show the usefulness of our method across a wide range of typical visualizations.
\end{abstract}



\begin{teaserfigure}
 	\centering
 \subfloat[Original]{\includegraphics[height = \teasorImgHeight]{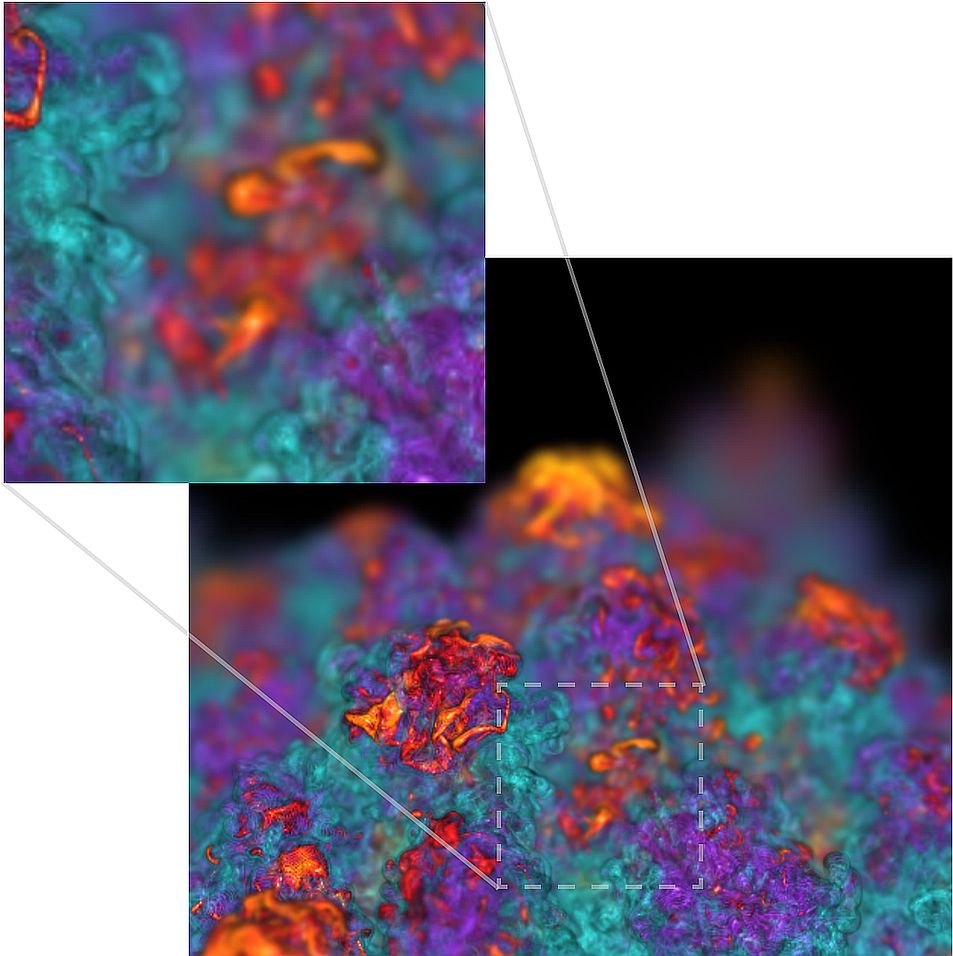}} 
 \hfill
 \subfloat[New method of visualization sharpening]{\includegraphics[height = \teasorImgHeight]{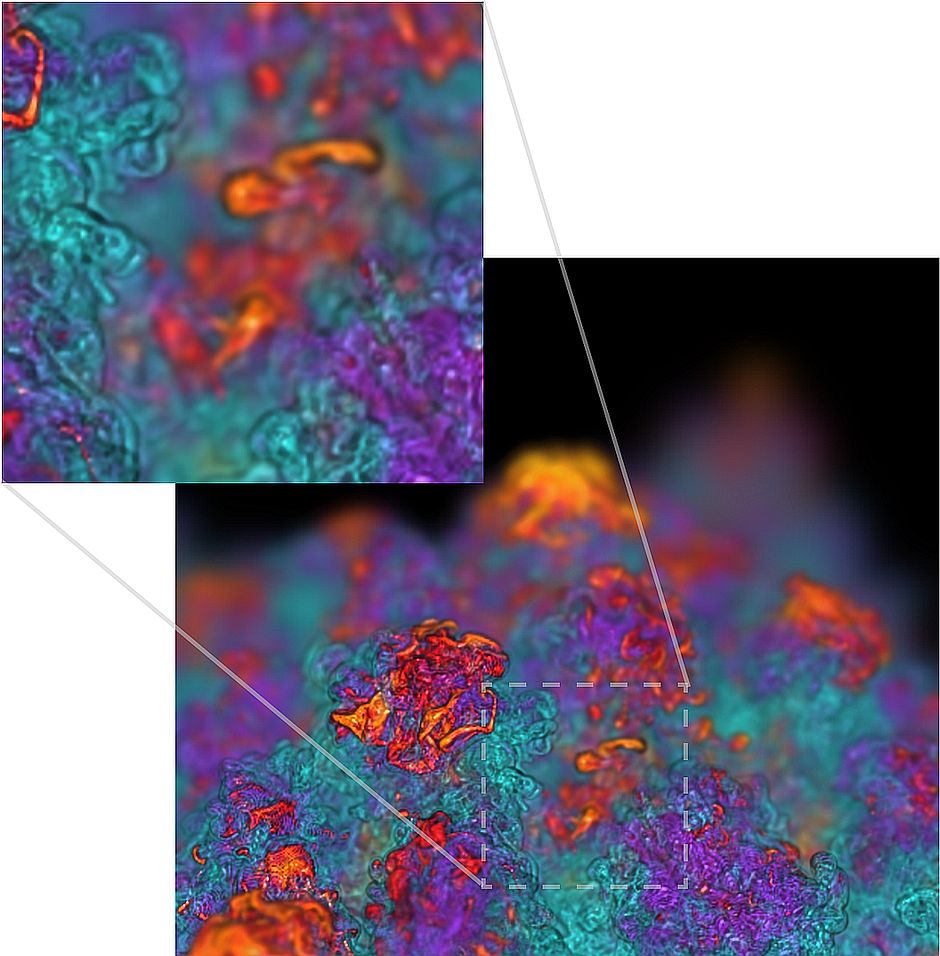}}
 \hfill
 \subfloat[Distance-based contrast enhancement~\cite{Zhou:VINCI} for comparison]{\includegraphics[height = \teasorImgHeight]{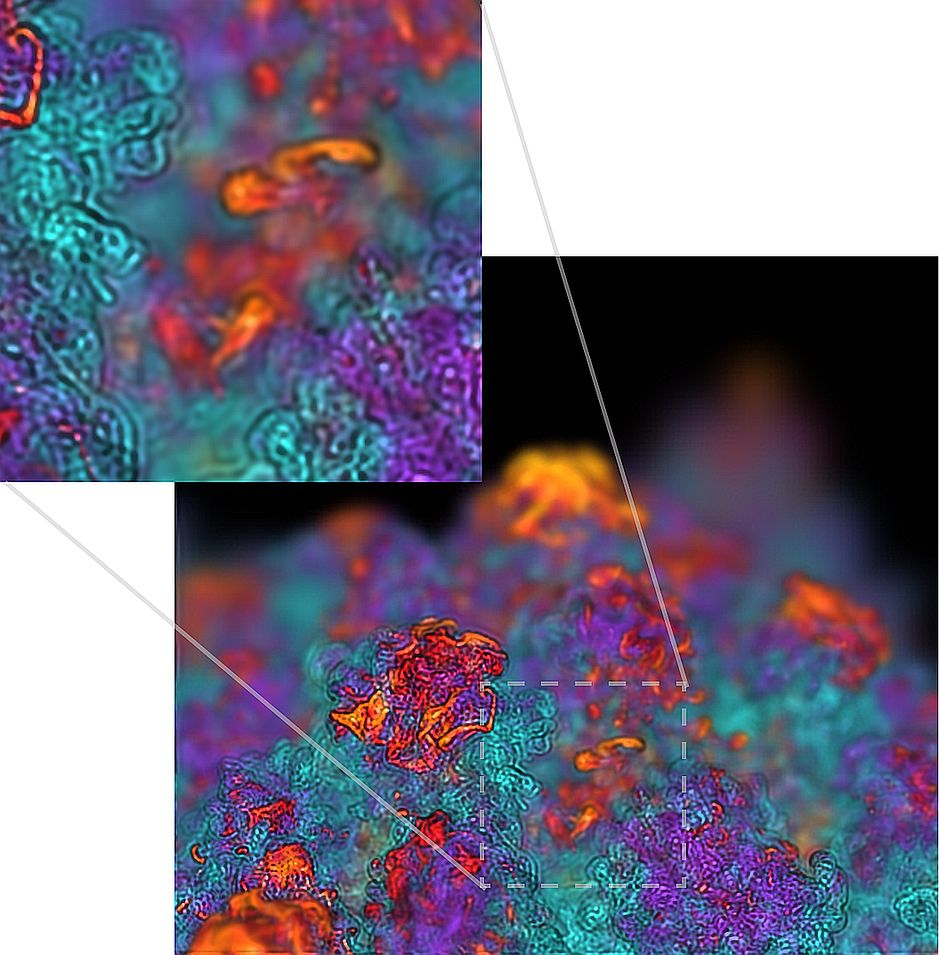}}
 \caption{A comparison of the original image (a) (\textcopyright\space John Wiley and Sons. Reprinted, with permission, from Schott et al.~\shortcite{Schott:2009:Eurovis}), spectral sharpened (b), and distance-based contrast enhanced~\cite{Zhou:VINCI} (c) on a volume rendering of a combustion simulation with depth-of-field effect. 
 	As shown in the zoom-ins, depth relationships of features can be better estimated with our method than the original, and our result is more natural-looking than the method of Zhou and Weiskopf~\shortcite{Zhou:VINCI}.
 	Note that figures are best viewed on a monitor as color reproduction of printers is not perfect.}\label{fig:introImg}
 \end{teaserfigure}

\maketitle
\section{Introduction}
To faithfully convey information through visualizations, the perception by the human recipient is equally important as the visual mapping pipeline that precedes it.
Contrast is essential in visual perception of color-coded visualizations---sufficient contrast is needed for showing boundaries of features.
In this paper, we perform spectral analysis of visualization perception under various viewing distances.
We propose a perceptually-guided multiscale method that sharpens visualizations by virtual viewing distance compensation.

Color mapping allows us to display data on a fine-grained level all the way down to per-pixel resolution, and it can convey both chromatic and achromatic information at the same time. 
It has been concluded that spatial frequency and contrast play important roles in the perception of chromatic and achromatic information of color-encoded visualizations~\cite{Ware:CGA:1988,Bergman:1995}.

Contrast sensitivity functions (CSFs) are an important tool to understand spatial vision. 
Researchers have measured CSFs in physiological and psychophysical experiments~\cite{VanNes:67, campbell1968application,wilson1991psychophysical,Mullen:JPhysio:1985}, and computational models of CSFs have been proposed~\cite{Movshon:88,barten1999contrast,Daly:92}.
Multiscale models~\cite{wilson1991psychophysical, Watson:97, PattanaikFFG98} can appropriately model spatial vision as the human visual system is believed to contain band-pass-fashioned visual pathways.
\rev{There is evidence that color CSFs behave differently from the luminance CSF, notably, color CSFs have peak values at lower spatial frequencies than the luminance CSF~\cite{Mullen:JPhysio:1985}, indicating that colors are more effective for encoding low-frequency features than luminance.}
CSFs have been used in visualization methods~\cite{Isenberg:VIS:2013,Zhou:VINCI} to enhance features of different scales.

Human visual perception is a complicated process, and CSFs are only concerned with threshold spatial vision, which predicts the visibility of an object under different viewing conditions.
To predict the appearances of objects that are visible, suprathreshold vision and spatial vision models have been studied~\cite{Georgeson:JPhysio:1975,Watson:97}.
Pattanaik et al.~\shortcite{PattanaikFFG98} propose a computational approach that realizes a comprehensive model that simulates perceptual phenomena in threshold/suprathreshold vision and apparent contrast under different illumination conditions.
This model serves as a basis for our new spectral sharpening method.

Our contribution is an image-based perceptually-inspired visualization sharpening technique.
We adopt the model of Pattanaik et al.~\shortcite{PattanaikFFG98}, and study the frequency domain of this model under various viewing distances and compensate for the power loss at a given viewing distance.
Specifically, the compensation is achieved by adapting the weights of band images of white noise data.
Therefore, our method implicitly accounts for perceptual effects beyond those described by CSFs.
An example can be seen in Figure~\ref{fig:introImg}, where depth-of-fielding volume renderings~\cite{Mathias:cgf11} are shown.
In the original rendering, depth relationships of out-of-focus features are difficult to judge (zoom-ins can be seen in insets). 
Using our method (Figure~\ref{fig:introImg}(b)), features in front are enhanced---one can conclude that they are closer to the focus than parts that are behind the focus. 
In comparison, the CSF-based approach~\cite{Zhou:VINCI} (Figure~\ref{fig:introImg}(c)) over-emphasizes high-frequency regions and causes ringing artifacts.

Our method decomposes an image into chromatic channels and a luminance channel comprised of multiscale bandpass images of the input image.
Chromatic channels are used to encode the main trend of the data that changes smoothly and has low spatial frequency, while luminance contrast is utilized for encoding small-scale value difference that exhibits high-frequency structures.
The effectiveness of our method is demonstrated by a wide range of visualization examples in the paper and the supplemental material.

Our approach has several advantages. One benefit is its generality---it works for visualizations with global illumination to 2D GIS examples. The method is an independent image processing approach that can be subsequently applied to any visualization system.
Our method is interactive, and overcompensation, which enhances features in visualizations, is achieved with easy-to-use user interaction---in the form of a single parameter of viewing distance. 

\begin{figure}[htb]
	\setlength{\tabcolsep}{2pt}
	\centering
	\begin{tabular}{ccc}
		\rot{90}{Input Images}	&
		\subfloat[]{\includegraphics[height = \HeightSimCompVis]{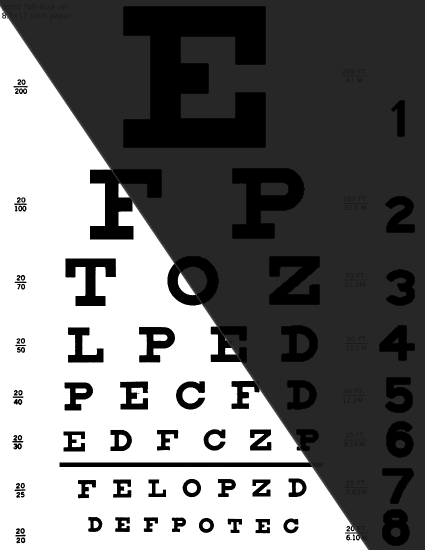}\label{fig:orgEyechart}}&
		\hfill
		\subfloat[]{\includegraphics[height = \HeightSimCompVis]{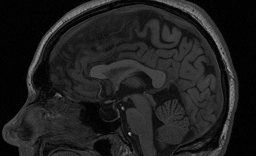}\label{fig:orgMRI}}
		\\
		
		\rot{90}{Viewing Distance 100\,cm}&
		\subfloat[]{\includegraphics[height = \HeightSimCompVis]{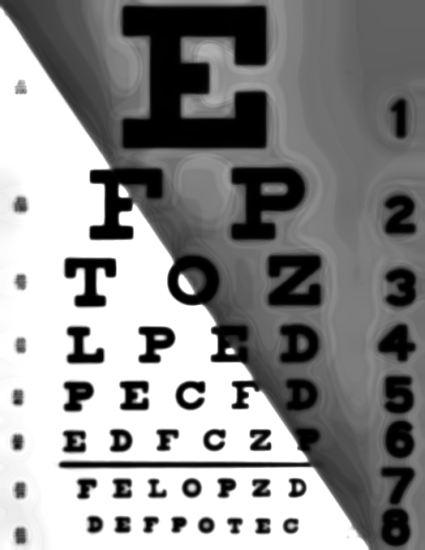}\label{fig:simEyechart}}&
		\hfill
		\subfloat[]{\includegraphics[height = \HeightSimCompVis]{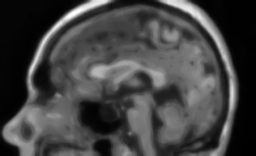}\label{fig:simMRI}}
		\\
		\rot{90}{Viewing Distance 10\,cm}&
		\subfloat[]{\includegraphics[height = \HeightSimCompVis]{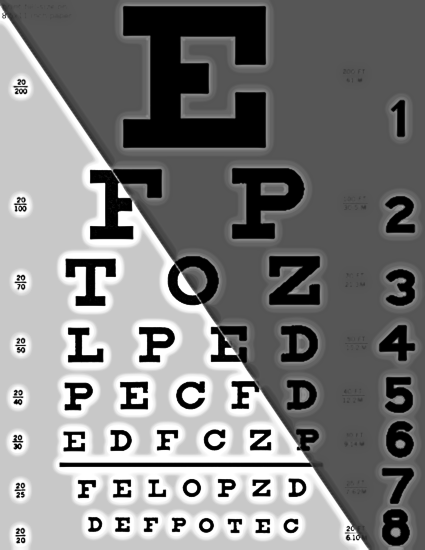}\label{fig:simEyechartClose}}&
		\hfill
		\subfloat[]{\includegraphics[height = \HeightSimCompVis]{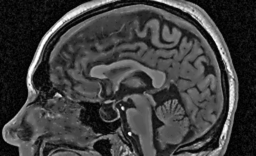}\label{fig:simMRIclose}}\\
	\end{tabular}
	\caption{
		Simulation results (c--f) from our implementation of the computational perception pipeline by Pattanaik et al.~\shortcite{PattanaikFFG98}. 
	}
	\vspace{-1em}
	\label{fig:percepPipeline}
\end{figure}

\section{Related Work}
\label{sec:RelatedWork}
Color mapping is an important visualization and perception research topic.
A survey on color mapping can be found elsewhere \cite{Zhou:TVCGPreprint}.
Among others, contrast, luminance, and spatial frequency are in particular related to our work. 
Luminance is found to be more effective for revealing high-spatial-frequency structures than chromatic channels~\cite{Ware:CGA:1988,Rogowitz:1996:LV:229737.229763}, which is in line with findings of psychophysical experiments~\cite{Mullen:JPhysio:1985}.
The combination of proper luminance and spatial frequency so that sufficient contrast can be perceived is important for successful color map design~\cite{Bergman:1995,Kovesionline}.
\rev{User studies~\cite{Padilla:VIS17} have shown that user task performance tends to be better with visualizations with discretized luminance than smooth luminance, implying that sufficient contrast is vital for effective visualization.
	Evidence of implicit discretization is found with chromatic channels as the result of an exploratory study of spectral color maps~\cite{Quinan:eurovis2019}.}

Contrast is also important in image processing and computer graphics as it is critical to improving image details.
Tone mapping operators~\cite{Reinhard:2002:PTR:566654.566575, Fattal:2002:GDH:566654.566573, Mai:TIP11:5648461} are concerned with the compression of luminance range while preserving perceived contrast. 
A perception-based tone mapping operator can simulate contrast reduction caused by glares in night driving~\cite{Meyer:2016:SVC}.
Unlike our proposed method, these image processing methods aim to reproduce the perceived image of high dynamic range input on low dynamic range displays, and cannot be tuned with a simple parameter.

An all-around and well-accepted computational perception model is by Pattanaik et al.~\shortcite{PattanaikFFG98}, which targets photorealistic image synthesis in computer graphics and---due to its broad coverage of perceptual phenomena \rev{such as threshold visibility, visual acuity, color discrimination, and suprathreshold brightness and colorfulness}---is useful for our visualization purposes; it unifies several important models from studies of the human visual perception.

We convert the resolution in this model from cycles per degree (cpd) to viewing distance, pixel count, and size. 
Then, we perform Fourier domain analysis on simulated images at various viewing distances and approximate the inverse of the model to enhance contrast by compensating for power loss.
Our aim is to enhance visualizations rather than simulating perceptual effects for accurate computer graphics rendering, and our approximated inversion leads to an efficient implementation of an interactive application that facilitates intuitive and easy-controllable user interactions.

\rev{Perceptual methods based on viewing distances and CSFs have been proposed in visualization.
	Isenberg et al.~\shortcite{Isenberg:VIS:2013} describe a multiscale visualization method for display walls by studying the visibility of features of different spatial frequencies at different viewing distances using CSFs and introduce a hybrid-image method for information visualization on a display wall by manually combining an image of high-frequency information and another of low-frequency information.}
Multiscale band-limited images are also used in our method, however, we utilize them for contrast enhancement and combine them automatically with adapted weights; unlike their power-wall setting, we focus on a typical working space setting with a fixed physical viewing distance from a regular monitor. 
Zhou and Weiskopf~\shortcite{Zhou:VINCI} propose calculating multiscale contrast for a given virtual viewing distance and test it against the threshold contrast curves---the inverse of CSFs---to enhance image bands that fall below the threshold contrast.
We also leverage the virtual viewing distance for intuitive and easy-controllable user interaction, but our model implicitly takes more perceptual effects into account as band weights are set based on the power spectral analysis of the computational perceptual simulation~\cite{PattanaikFFG98}.
As a result, a more balanced weight combination is achieved in our method---yielding more natural-looking results than the previous method~\cite{Zhou:VINCI}, which potentially amplifies high frequencies too much when lower frequencies are not amplified at all. 
Furthermore, our spectral contrast model can take the non-colormapped original data as input, which provides additional luminance details in the enhanced results as shown in Figures~\ref{fig:introImg},~\ref{fig:MethodOverview}, and~\ref{fig:sliceResults}.

Distance perception is critical in virtual reality environments and has been extensively studied~\cite{Interrante:VR06,Vaziri:2017:IVE}.
Although we focus on the use of virtual viewing distance as leverage for contrast enhancement, it is possible to extend our work to virtual and augmented reality settings. 

\begin{figure}[htb]
	\begin{center}
		\includegraphics[width=\linewidth]{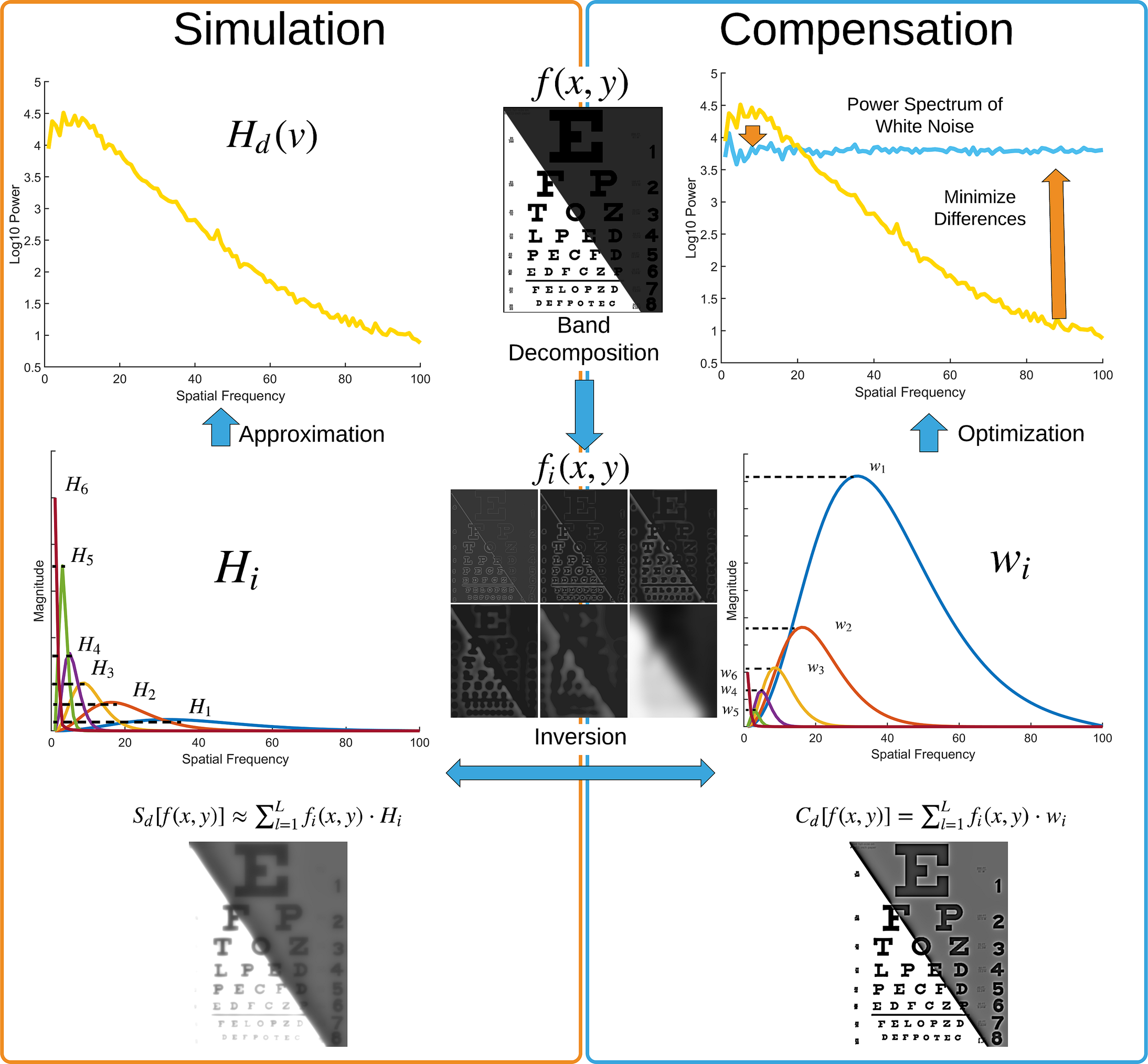}		
		\caption{The spectral perceptual model (Section~\ref{sec:specPercpModel}) is illustrated on the left, and the compensation pipeline (Section~\ref{sec:weightAssign}) is shown to the right.}
		\vspace{-1em}
		\label{fig:model}
	\end{center}
\end{figure}

\section{Spectral Visualization Sharpening}
\label{sec:lessonsPercepModel}
We propose a spectral image sharpening method based on Fourier domain analysis of the computational perceptual simulation model by Pattanaik et al.~\shortcite{PattanaikFFG98} that generates convincing perceptual images.
Intuitively, the goal of compensation for contrast loss due to viewing distance is to make an image appear identical to the original image at this viewing distance. 
Therefore, we need to approximate the inverse of the perceptual simulation.
Given the evidence that the human visual system behaves as spatial-frequency filters, it is appropriate to investigate the effect of the perceptual pipeline by analyzing in the Fourier domain.
In this way, we build a simplified model to achieve the compensation based on the power spectrum of an image as shown in Figure~\ref{fig:model}~(left).

This model allows us to find the inversion of the simulation, i.e., to compensate for contrast loss, by controlling weights of band-limited images (Figure~\ref{fig:model}~(right)).
The compensation is formulated as an optimization problem (Section~\ref{sec:weightAssign}).
Furthermore, the model enables us to reduce computational complexity and achieve an interactive method that can be added into any interactive visualization system.

\subsection{Simulated Impact of Viewing Distance}
\label{sec:visionModel}
We study the impact of viewing distance as a perceptual parameter in the model of Pattanaik et al.~\shortcite{PattanaikFFG98} by converting its resolution measurement from cpd to viewing distance $d$ and pixel measurements~\cite{Isenberg:VIS:2013}.
Since our method preserves chromatic channels and sharpens the luminance channel, the input to the pipeline is a grayscale version of the original image\rev{---we calculate the luminance $Y$ in $XYZ$ color space from the linear RGB input}.

Results from our implementation are shown in Figure~\ref{fig:percepPipeline} with test images: Snellen eye chart (a test image also used in the original publication of the pipeline~\cite{PattanaikFFG98}) and a slice through an MRI brain scan  (as a typical example from scientific and medical visualization).
The physical sizes for the images are 6.9\,cm~$\times$ 8.9\,cm, 4.1\,cm $\times$ 2.5\,cm, and 12.3\,cm $\times$ 7.6\,cm, respectively, on a 28-inch monitor with a resolution of 3840 $\times$ 2160 pixels.

Several qualitative characteristics should be noted in the simulation results. 
Compared to the input images (Figures~\ref{fig:orgEyechart}--\ref{fig:orgMRI}), results with $d= 100$\,cm are much more blurry and fine details are lost, e.g., the small numbers in the Snellen chart (Figure~\ref{fig:simEyechart}) and the details of the cortex structure  (Figure~\ref{fig:simMRI}). 
Comparing results with short viewing distance (Figures~\ref{fig:simEyechartClose} and~\ref{fig:simMRIclose}) to long viewing distance,   
the adaptation becomes more local with a shorter viewing distance, e.g., the halo effects in the Snellen chart with $d=10$\,cm make the image much crisper compared to the $d=100$\,cm version, the boundaries are enhanced, and the contrast of images is increased.

Overall, the perceptual simulation tends to blur the original images for medium to large viewing distances. Therefore, we do not perceive visual patterns at small length scales as well as they are in the original data. 

\subsection{Fourier Analysis of Perception Simulations}
\label{sec:powerSpecPercepModel}
To understand the frequency behavior of the perceptual pipeline on visualizations, we conduct Fourier analysis---using the radial power spectrum operator $Pr[\cdot](v)$, where $v$ is the radial frequency---on simulated perceptual images (full details of the analysis and images of visualizations are documented in the supplemental material). 
We analyze 50 visualizations obtained through Google image search---these images cover typical classes of visualizations, including visualizations of volume, flow, DTI, GIS data, and slices of medical scans. 
Each image is simulated with viewing distances from 10\,cm to 100\,cm with a stride of 10\,cm, and we calculate the logarithmic power spectrum of each original image and its simulations.\rev{ Perceived changes over spatial frequency of stimuli behave roughly logarithmically according to the Weber-Fechner law}.
We are interested in the relative relationship between the power spectrum of the original image and simulated images.

\begin{figure}[htb]
\vspace*{-1em}
	\centering
	\subfloat[Averaged Relative Power Spectra of 50 Visualizations]{\includegraphics[width=0.75\linewidth]{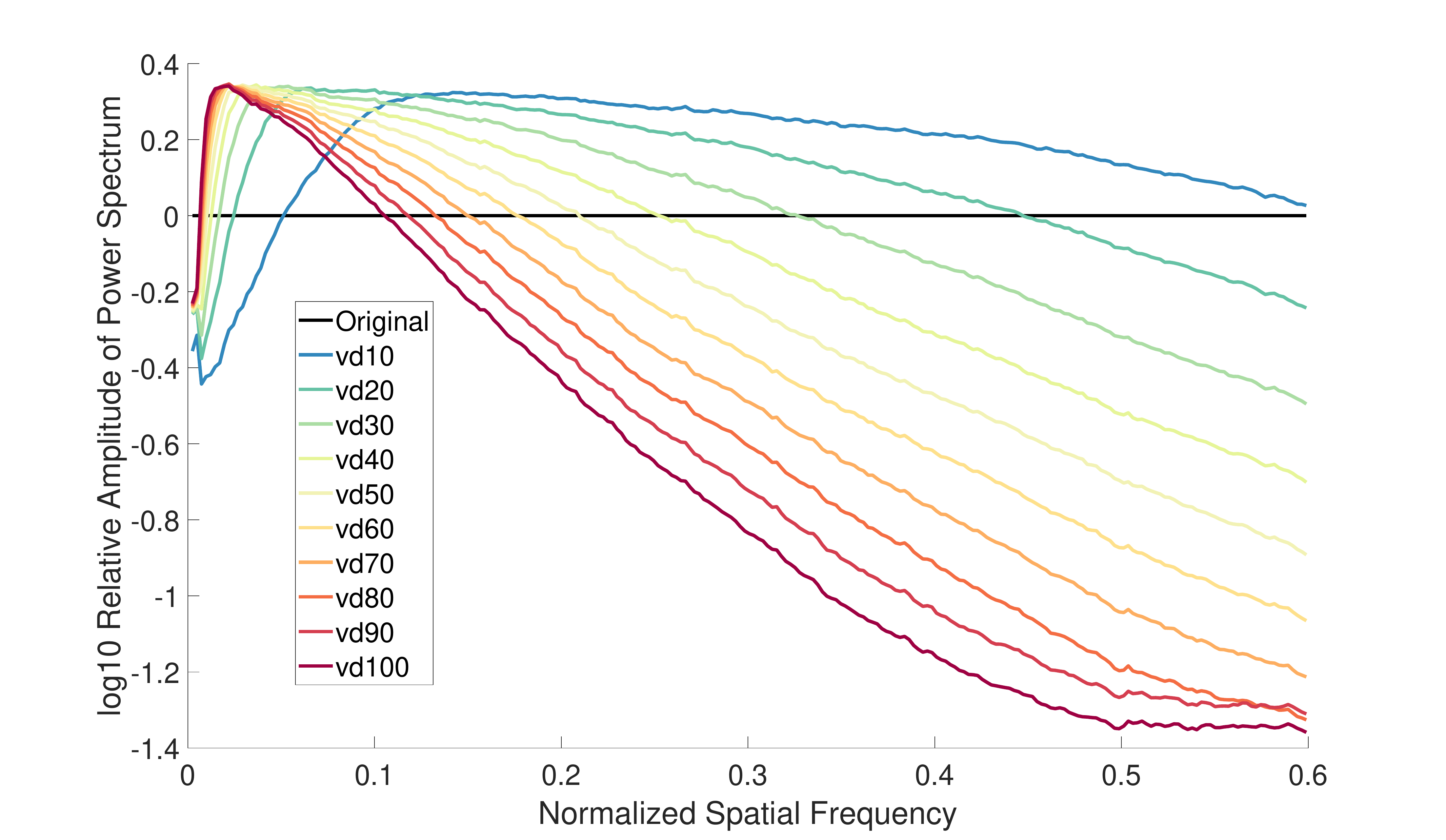}\label{fig:avgSpecRel}}
	\\
	\vspace{-1em}
	\subfloat[Relative Power Spectra of White Noise]{\includegraphics[width=0.75\linewidth]{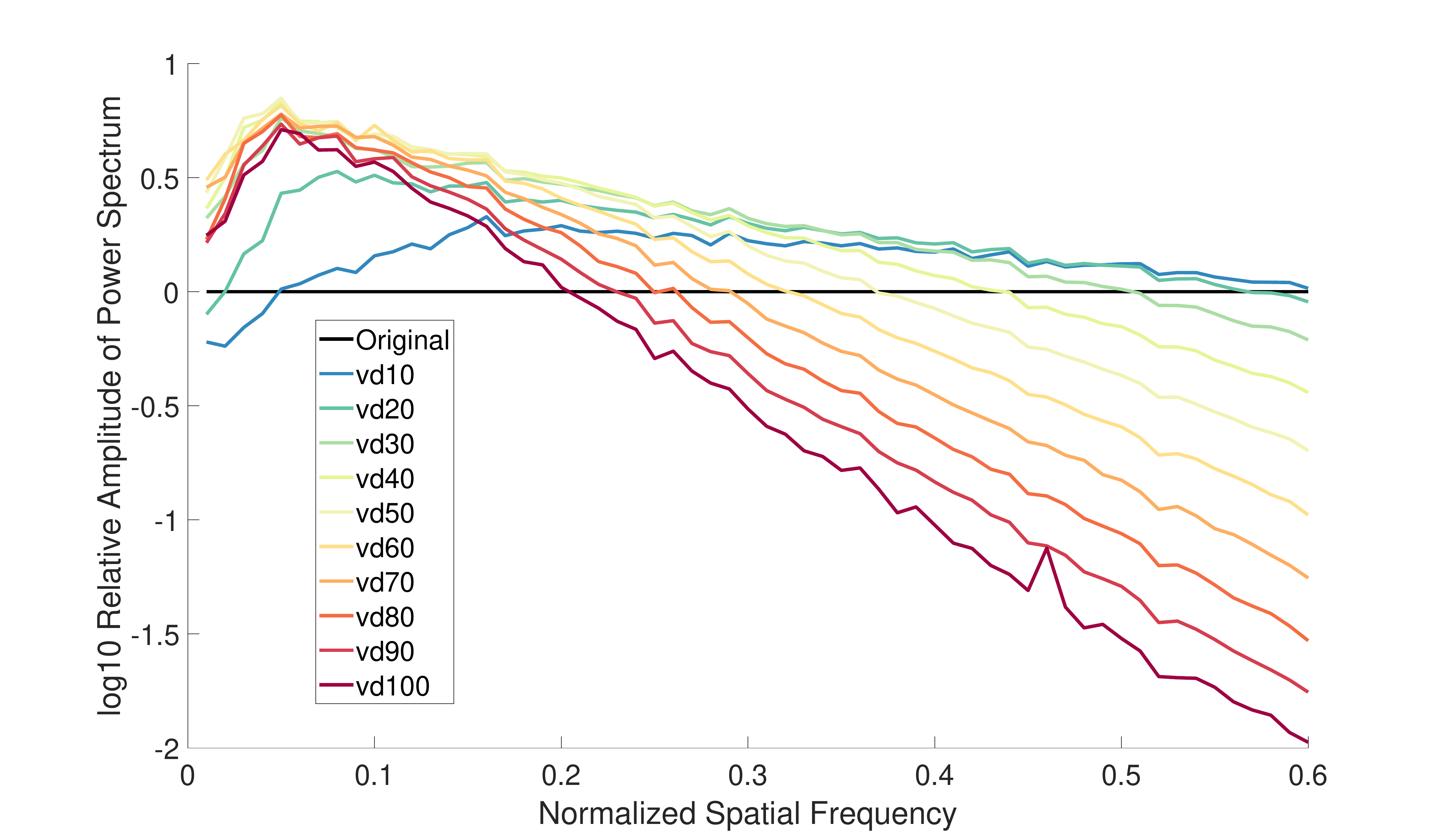}\label{fig:whitenoiseRel}}\\
	\vspace{-1em}
	\caption{Logarithmic relative amplitude of power spectra of simulated perceptual images of 50 visualizations (a), and white noise (b). Here, we use the normalized spatial frequency.}
	\label{fig:allPowerSpec}
\end{figure}
Therefore, we derive the relative amplitude by dividing the power spectra of simulations by the spectrum of the original image, and aggregate all datasets and calculate the mean relative amplitude of each viewing distance.
The result is shown in Figure~\ref{fig:allPowerSpec}(a), where the spatial frequency is limited to roughly 60\% of the averaged highest frequency of all images as the very high frequency is not reliable against artifacts. 
From these power spectra, we can observe: the perceptual pipeline mostly behaves like a bandpass filter, and the filter response works in a controlled way, i.e., without rapid changes; the power spectrum of an image simulated with a greater viewing distance decays more quickly compared to an image with a shorter viewing distance; power spectra of long viewing distances are higher than the original in low-middle frequencies and lower than the original in high frequencies, and those of short viewing distance behave the opposite. 

\begin{table}[htb]
	\centering
	\caption{Slopes from linear regression of relative power spectra of visualization images (vis) and white noise (wn) simulations of $d=$10\,cm to 100\,cm. }
	\label{tab:linReg}
	\begin{tabular}{ccc}
		\toprule
		$d$ & \textbf{vis} (Figure~\ref{fig:allPowerSpec}(a)) & \textbf{wn} (Figure~\ref{fig:allPowerSpec}(b))\\\midrule
		10 & -0.31 & -0.44\\
		20 & -0.82 & -1.02\\
		30 & -1.40 & -1.59\\
		40 & -1.99 & -2.20\\
		50 & -2.51 & -2.80\\
		60 & -2.94 & -3.37\\
		70 & -3.28 & -3.88\\
		80 & -3.59 & -4.36\\
		90 & -3.87 & -4.78\\
		100 & -4.16 & -5.14\\\bottomrule
	\end{tabular}
\end{table}
To model the spectral response of the perceptual pipeline in a data-independent way (we refer to data in the visualization images), we study the frequency behavior of the pipeline for gradually increasing viewing distances using a white noise image, where samples are randomly drawn from a Gaussian probability distribution.
\begin{figure*}[tb]
	\begin{center}
		\includegraphics[width = 0.85\linewidth]{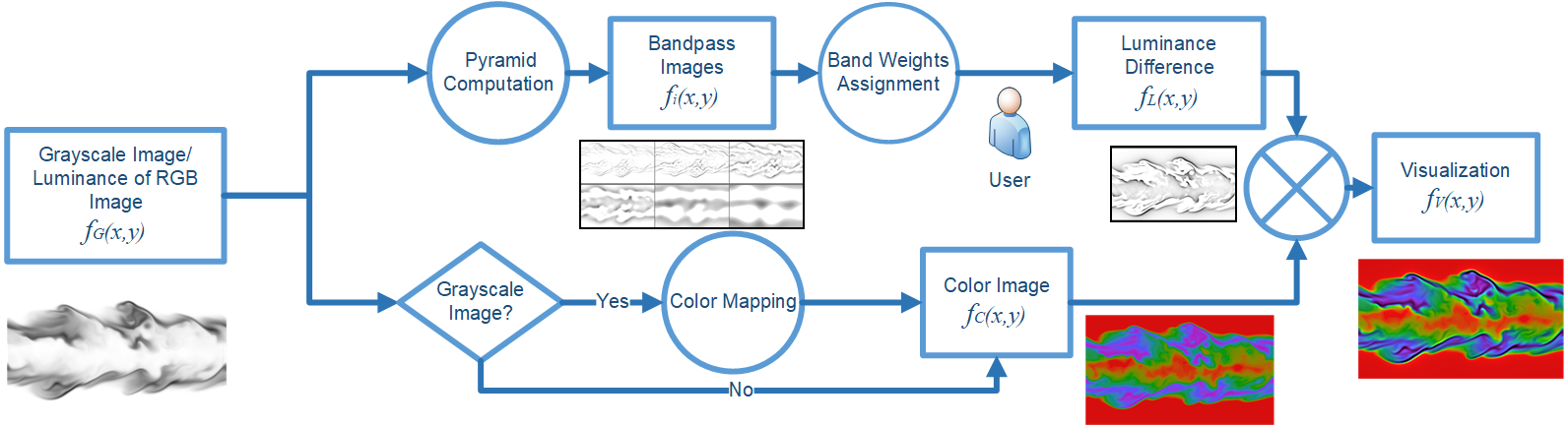}
		\caption{Flow chart of the proposed method. Luminance images $f_i(x,y)$ and $f_L(x,y)$ are inverted for visibility.}
		\label{fig:MethodOverview}
		\vspace{-1em}
	\end{center}
\end{figure*}
The power spectrum of a white noise image is constant across all frequencies in the Fourier domain---up to small variations from the stochastic construction.
Therefore, white noise provides a good means of assessing how the perception pipeline affects various spatial frequencies, avoiding any potential artifacts from regular sampling.
Figure~\ref{fig:allPowerSpec}(b)---plotted in normalized spatial frequency, and we limit the highest frequency to 60\% to match Figure~\ref{fig:allPowerSpec}(a)---shows relative amplitude of logarithmic power spectra 
of a white noise image and its simulations from the perception pipeline of $d=10$\,cm through 100\,cm with a stride of 10\,cm.
Compared to the relative amplitude of real datasets (Figure~\ref{fig:allPowerSpec}(a)) in the middle range of frequency, i.e., the normalized spatial frequency of 0.1---0.35, the white noise curves behave qualitatively similar and quantitatively comparable.
Linear regression is calculated for both relative power spectra of visualizations (spatial frequency: 0.1---0.35) and white noise (spatial frequency: 0.1---0.6), and the slopes of the fitted lines are shown in Table~\ref{tab:linReg}.

Therefore, it is valid to use white noise as a representative for the power spectra of visualization images.

\subsection{Spectral Perceptual Model}
\label{sec:specPercpModel}
We have a filtering that is frequency-dependent, and therefore, we formulate a spectral perception model for an image $f(x,y)$:
\begin{equation}
S_d[f(x,y)] = \mathscr{F}^{-1}[\mathscr{F}[f(x,y)] \cdot H_{d}(\nu)] \;,
\label{eqn:simpModel}
\end{equation}
where $S_d[\cdot]$ is the perceptual simulation operation for a virtual viewing distance $d$, and $H_d$ is a transfer function of radial frequency $v$. 
As shown in Equation~\ref{eqn:transfunc}, $H_d$ is the radial power spectrum of the white noise data simulated at $d$ (one of the curves in Figure~\ref{fig:allPowerSpec}(b)):
\begin{align}
H_d(\nu) = P_r[\mathscr{F}[S_d[n(x,y)]]](\nu)\;.
\label{eqn:transfunc}
\end{align}
Here, we assume that the original noise image $n(x,y)$ is normalized in the sense that its power spectrum averages to one.

We can replace the input image by the sum of $L$ images in a Gaussian pyramid; therefore, Equation~\ref{eqn:simpModel} can be rewritten as:
\begin{align}
S_d[f(x,y)] &= \mathscr{F}^{-1}\left[\sum_{i=1}^{L}\mathscr{F}[f_i(x,y)] \cdot H_{d}(\nu)\right]\;.
\end{align}
Since the transfer function changes in a controlled way, it is valid to approximate the transfer function on each frequency interval of the bandpass images with a constant $H_i$.
The Fourier transform is a linear operator, therefore we have:
\begin{align}
\centering
S_d[f(x,y)] &\approx \mathscr{F}^{-1}\left[\sum_{i=1}^{L}\mathscr{F}[f_i(x,y)] \cdot H_i\right] \\
&= \sum_{i=1}^{L}H_i\cdot f_i(x,y)\label{eqn:simpModelApprox} \;.
\end{align}
This equation serves as the mathematical model of our simplified perception pipeline.
Figure~\ref{fig:model} illustrates the spectral perceptual model on the left.

It is possible to compute the inverse of the model for a virtual viewing distance $d$, by replacing the constants $H_i$ with their inverse $1/H_i$ in Equation~\ref{eqn:simpModelApprox}. 
Effectively, it raises the power spectrum of the perceived image at $d$ to the constant value, leading to perceptual compensation.

\subsection{Spectral Visualization Sharpening Pipeline}
The basis of our spectral sharpening pipeline (Figure~\ref{fig:MethodOverview}) is reverting the frequency transfer of Equation~\ref{eqn:simpModelApprox} to invert the frequency damping of the image.
Given an input image $f(x,y)$ in its gray-scale version $f_G(x,y)$, a number of $L-1$ band images $f_i(x,y)$ are derived by taking the top level and then calculating the differences between two neighboring levels in a Gaussian pyramid.
It is then possible to compensate for contrast loss via the power spectrum by assigning proper weights to band images.
In Section~\ref{sec:weightAssign}, the weight optimization method for virtual viewing distance compensation is explained.

Finally, a visualization $f_V(x,y)$ is generated by combining the luminance difference image $f_L(x,y)$, which is a weighted sum of bandpass images $f_i(x,y)$, and the color image $f_C(x,y)$. 
Here, the $f_C(x,y)$ contains the full chromatic information and the lowpass version of the achromatic image.
Therefore, the achromatic channel of the final image is the sum of the achromatic channel of $f_C(x,y)$ and $f_L(x,y)$, the chromatic channels are directly taken from $f_C(x,y)$.

\begin{figure*}[tb]
	\centering
	\begin{tabular}{cccc}
		\subfloat[Original]{\includegraphics[height = \tomatoHeight]{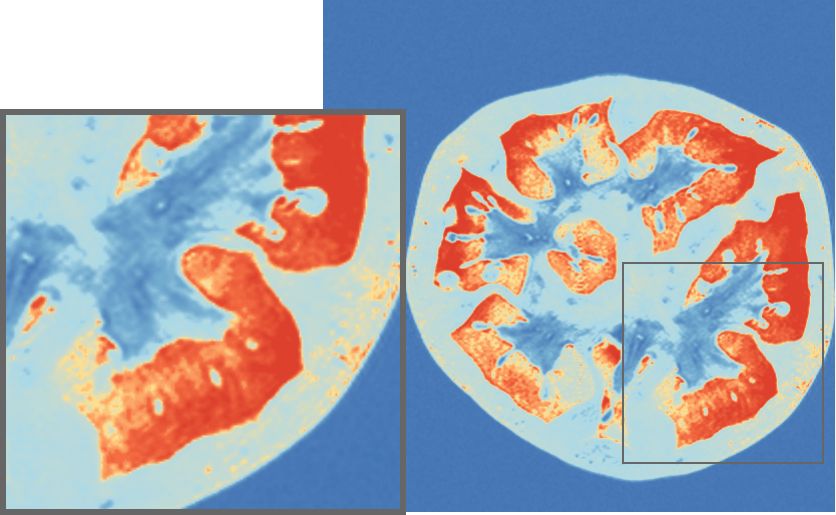}\label{fig:tomatoCm}} &
		\subfloat[Compensated]{\includegraphics[height = \tomatoHeight]{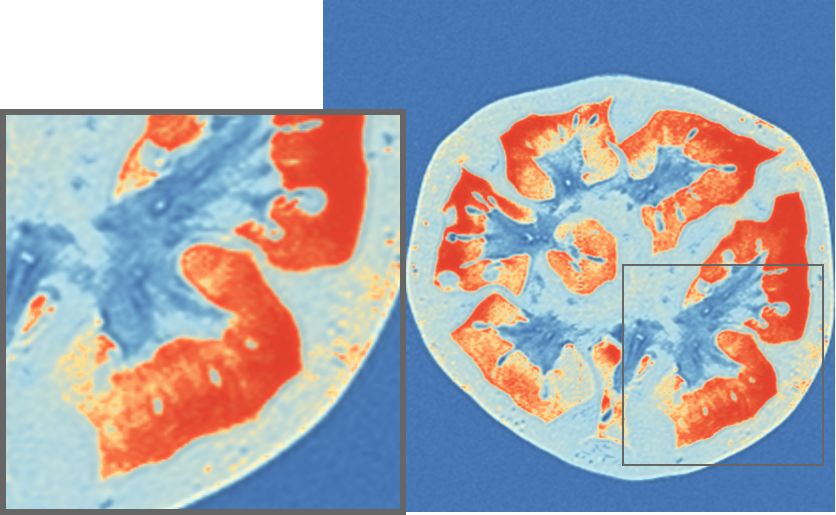}\label{fig:tomatoVd60}}&
		\subfloat[Overcompensated]{\includegraphics[height = \tomatoHeight]{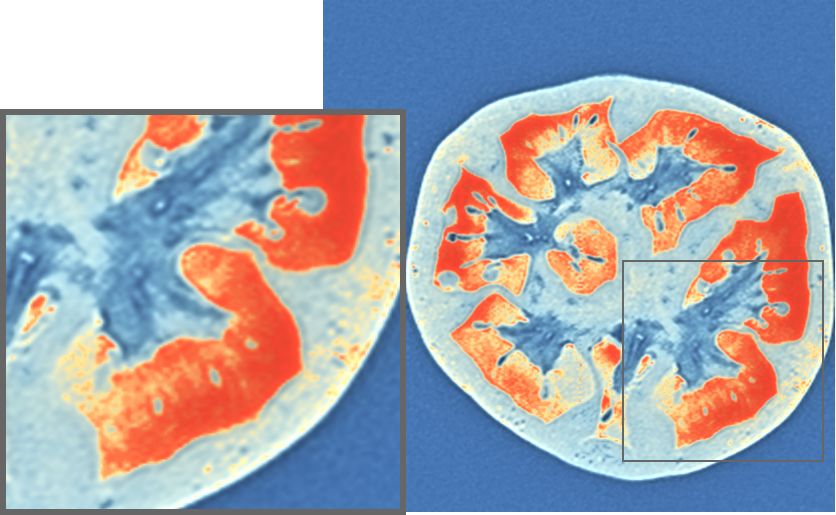}\label{fig:tomatoCmlumi}}&
		\subfloat{\includegraphics[height = \tomatoHeight]{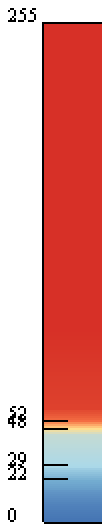}\label{fig:tomatoColormap}}
	\end{tabular}
	\caption{A comparison of no compensation (a), compensation (b), and overcompensation (c) for contrast on a slice of a CT scan of a tomato. }\label{fig:tomato}
	\vspace{-1em}
\end{figure*}
\section{Compensation}
\label{sec:weightAssign}

In this section, we explain compensation---the formulation of the problem and its solution (Figure~\ref{fig:model}~right)---and overcompensation.

\subsection{Mathematical Formulation for Compensation}
Consider the compensation for a virtual viewing distance $d$ as an operator $C_d[\cdot]$, along with the simulation operator $S_d[\cdot]$ for our simplified perceptual model. We can then formulate the compensation problem as follows:
\begin{align}
&S_d[C_d[f(x,y)]] \buildrel !\over= f(x,y) \label{eqn:compensation}\;.
\end{align}
Equation~\ref{eqn:compensation} essentially says that the simulated result from our simplified perception model of a compensated image should be equal to the input image $f(x,y)$.
In exactly the same way of $S_d[\cdot]$, the compensation operator has the following form:
\begin{align}
&C_d[f(x,y)] = \sum_{i=1}^{L}w_{i}\cdot f_{i}(x,y)\text{,\space and\space}w_i \geq 0\;,
\end{align} 
where $w_i$ are non-negative weights for bandpass images.
We do not modify the lowpass image, i.e., $w_L = 1$. 

\subsection{Finding Optimal Weights}
In order to derive a set of $w_i$ in a data-independent fashion, we apply the white noise image $f(x,y) \equiv n(x,y)$, and apply the Fourier transform to both sides of Equation~\ref{eqn:compensation}:
\begin{align}
\mathscr{F}[S_d[C_d[n(x,y)]]] &\buildrel!\over= \mathscr{F}[n(x,y)]~\label{eqn:Fboth}\\
\Rightarrow\sum_{i=1}^{L}w_i\mathscr{F}[S_d[n_i(x,y)]] &\buildrel !\over= \sum_{i=1}^{L}\mathscr{F}[n_i(x,y)] \label{eqn:bandCompensation}
\end{align}
It is possible to approximate Equation~\ref{eqn:bandCompensation} by minimizing the difference between its left- and right-hand sides: 
\begin{align}
\arg\!\min_{w_i}\text{\space}p\quad, \text{\space subject to\space}w_i \geq 0\;,
\label{eqn:optForm}
\end{align}
with the objective function:
\begin{align}
p = \left\lVert\sum_{i=1}^{L}\{w_i\mathscr{F}[S_d[n_i(x,y)]] - \mathscr{F}[n_i(x,y)]\}\right\rVert_2\;.
\end{align}

The objective function is then simplified: we drop the phase information of the Fourier transform, and evaluate only the power spectrum $P_r[\cdot]$.
Also, instead of evaluating the power spectrum for the whole frequency range, the band-limited power spectrum of the middle frequency is used for its robustness against artifacts. 
Therefore, the objective function $p$ reads:
\begin{align}
p=\int_{a\nu_{m}}^{b\nu_{m}}\Biggl(\sum_{i=1}^{L}\{w_{i}&P_r[\mathscr{F}[S_d[n_i(x,y)]]](\nu)-
\notag\\
&P_r[\mathscr{F}[n_i(x,y)]](\nu)\}^2\Biggr)\mathrm{d}v\text{,\space }w_i \geq 0\;,
\end{align}
where $\nu_{m}$ is the maximum frequency of the image, and $a$ and $b$ are values satisfying $0 < a < b < 1$; $a = 0.05$ and $b=0.6$ are utilized in our implementation. 
Recalling Figure~\ref{fig:allPowerSpec}, $P_r[\mathscr{F}[S_d[n_i(x,y)]]](\nu)$ is essentially the band-limited power spectrum  of the white noise image simulated at a distance $d$, whereas $P_r[\mathscr{F}[n_i(x,y)]](\nu)$ is the constant power spectrum of the original image.

Finally, the optimal weights are found by solving a constrained nonlinear optimization
of Equation~\ref{eqn:optForm}.
We solve this problem using the conjugate gradient method~\cite{nocedal2006numerical}, which has good convergence performance. 
Since it is impossible to analytically compute the gradient of $p$, we approximate the gradient with central differences. 

We only need to compute the optimal weights for a white noise image, and therefore the optimization is conducted in a preprocessing stage.
Optimal weights $w_i$ found through the process are stored and loaded at runtime during the visualization.

\subsection{Overcompensation}
\label{sec:overcomp}

In many cases, it is necessary to emphasize details in the image for visualization purposes rather than just to compensate for contrast loss.
Such overcompensation can be easily achieved by setting the virtual viewing distance parameter $d$ to be greater than the actual viewing distance. 
Comparing the result of overcompensation as in Figure~\ref{fig:tomato}(c) to compensation in Figure~\ref{fig:tomato}(b), it is noticeable that overcompensation makes structure boundaries have increased contrast and details become more visible.
Specifically, overcompensation improves the visibility of details in the placental tissue in blue, regions on the pericarp wall colored in light blue, and also adds halo effects for the boundary of the tomato.
\begin{figure*}[htb]
	\setlength{\tabcolsep}{2pt}
	\centering
	\begin{tabular}{cccc}
		&
		Color-mapped data &
		Our method &
		Color map\\
		\rot{90}{MRI brain scan}    &
		\subfloat{\includegraphics[width = \volDataWidthCm]{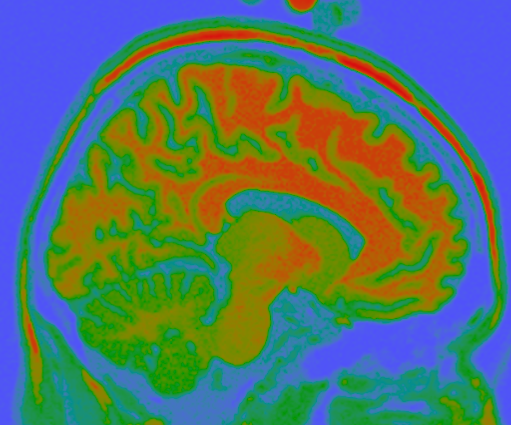}\label{fig:MRIbrainCm}} &
		\subfloat{\includegraphics[width = \volDataWidthCm]{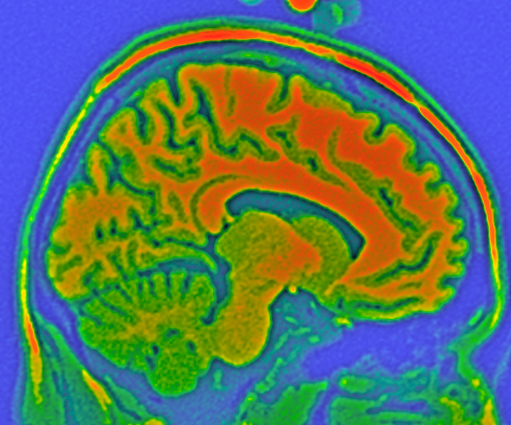}\label{fig:MRIbrainCmlumi}}&
		\subfloat{\includegraphics[height = \volDataHeight]{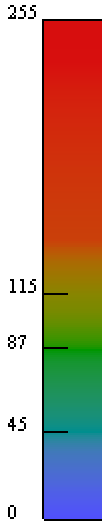}\label{fig:MRIColormap}}\\
		\rot{90}{Hurricane Isabel pressure}&
		\subfloat{\includegraphics[width = \volDataWidthCm]{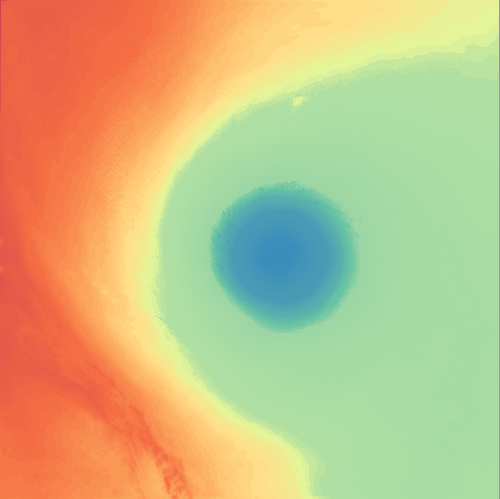}\label{fig:isabelCm}} &
		\subfloat{\includegraphics[width = \volDataWidthCm]{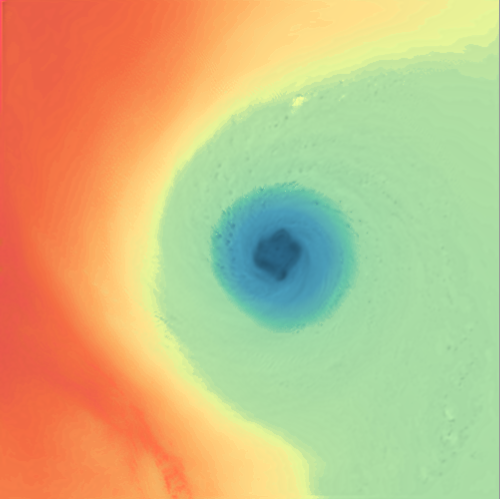}\label{fig:isabelCmlumi}}&
		\subfloat{\includegraphics[height = \volDataHeight]{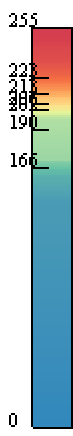}\label{fig:isabelPcolormap}}
	\end{tabular}
	\caption{Examples of volumetric data slices. The virtual viewing distance used to create results with our method are 75\,cm for the MRI brain data set, and 90\,cm for the Hurricane Isabel pressure dataset.
	}
	\label{fig:sliceResults}
	\vspace{-1em}
\end{figure*}

\rev{
	The user can interactively change the virtual viewing distance by controlling a single slider. 
	In our GPU-accelerated implementation, the optimal band weights $w_i$ are precomputed and are used during runtime to compute a luminance image, which further composites with the chromatic image.
}

\section{Examples}
\label{sec:results}
We show the usefulness of our sharpening method for a wide range of typical applications: color mapping on 2D images (slices) to show scalar fields, 3D volume renderings, and 2D map-based geographic information visualization. 
A video of screen captures of interactions with these datasets can be found in the supplemental material.

\begin{figure*}[htb]
	\setlength{\tabcolsep}{2pt}    
	\centering
	\begin{tabular}{ccc}
		&
		Original image &
		Our method \\
		\rot{90}{Volume scattering (vsc)}&
		\subfloat{\includegraphics[width = \imgDataWidth]{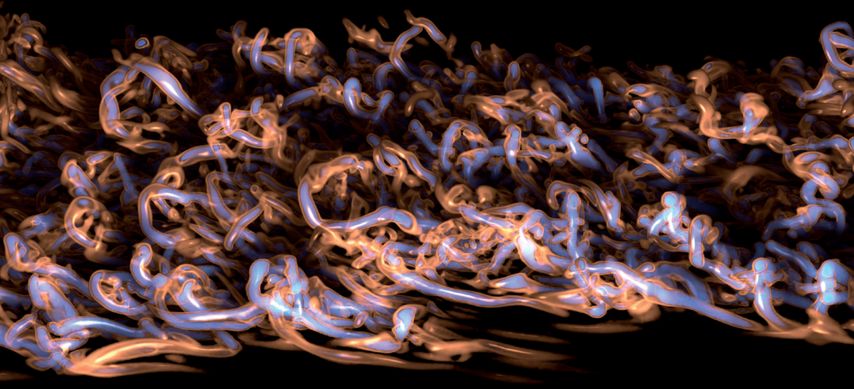}\label{fig:vortex_vis13}} &
		\subfloat{\includegraphics[width = \imgDataWidth]{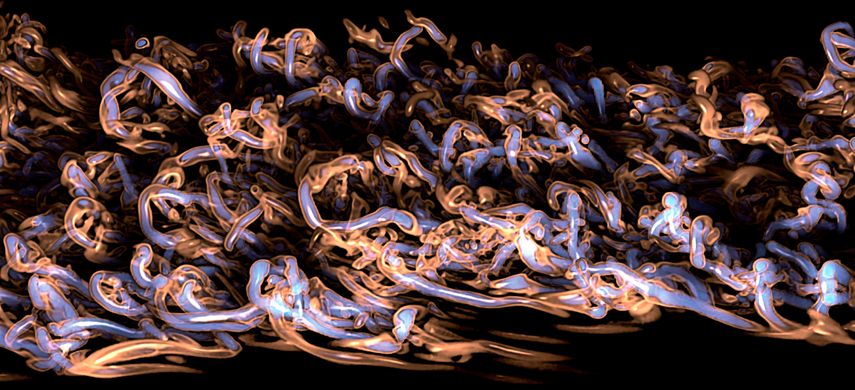}\label{fig:vortex_vis13_enhanced}} \\
		\rot{90}{Volume shadowing (vsh)}    &
		\subfloat{\includegraphics[width = \imgDataWidth]{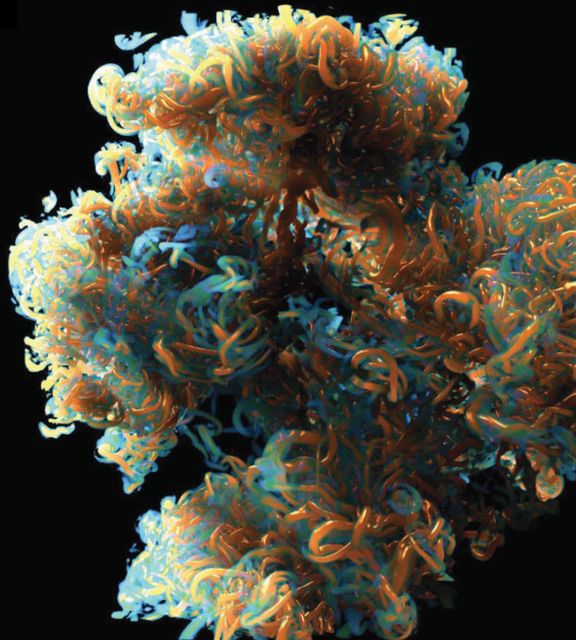}\label{fig:vortex_vis14}} &
		\subfloat{\includegraphics[width = \imgDataWidth]{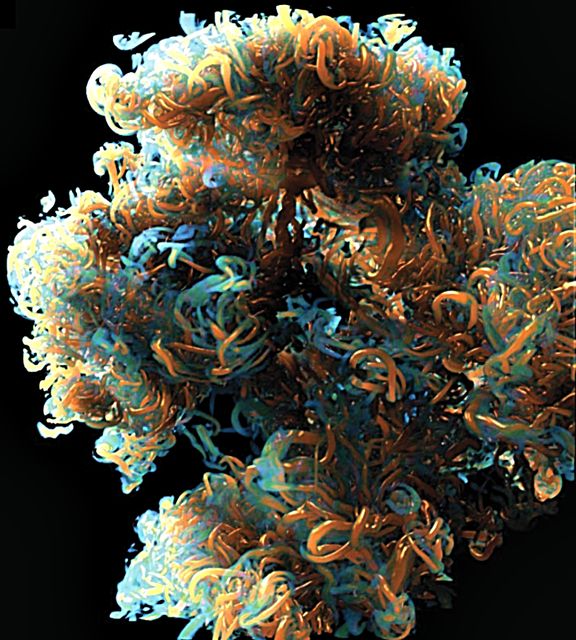}\label{fig:vortex_vis14_enhanced}}\\
		\rot{90}{GIS visualization (gis)}&
		\subfloat{\includegraphics[width = \imgDataWidth]{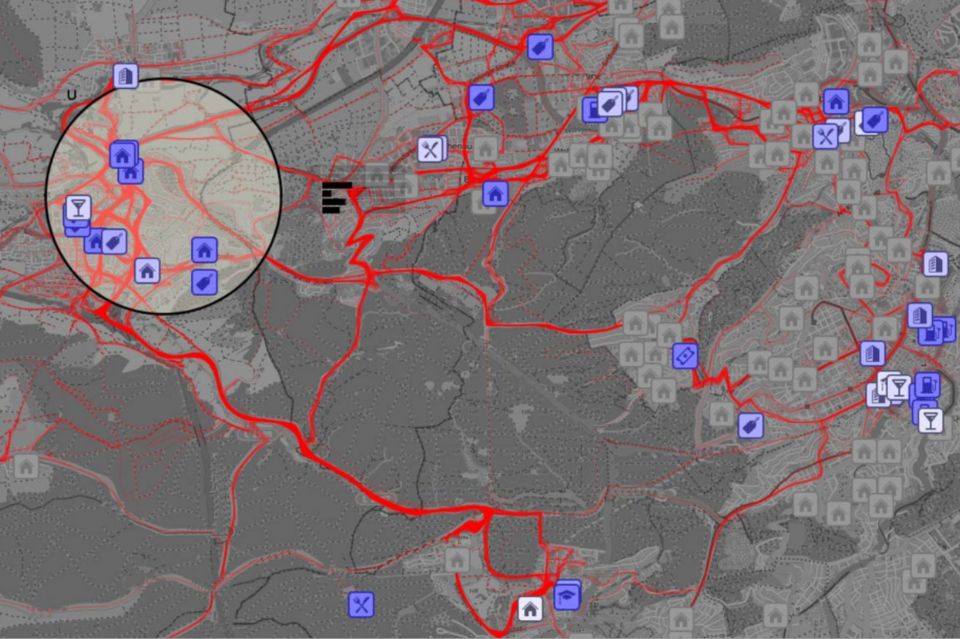}\label{fig:GIS}}&
		\subfloat{\includegraphics[width = \imgDataWidth]{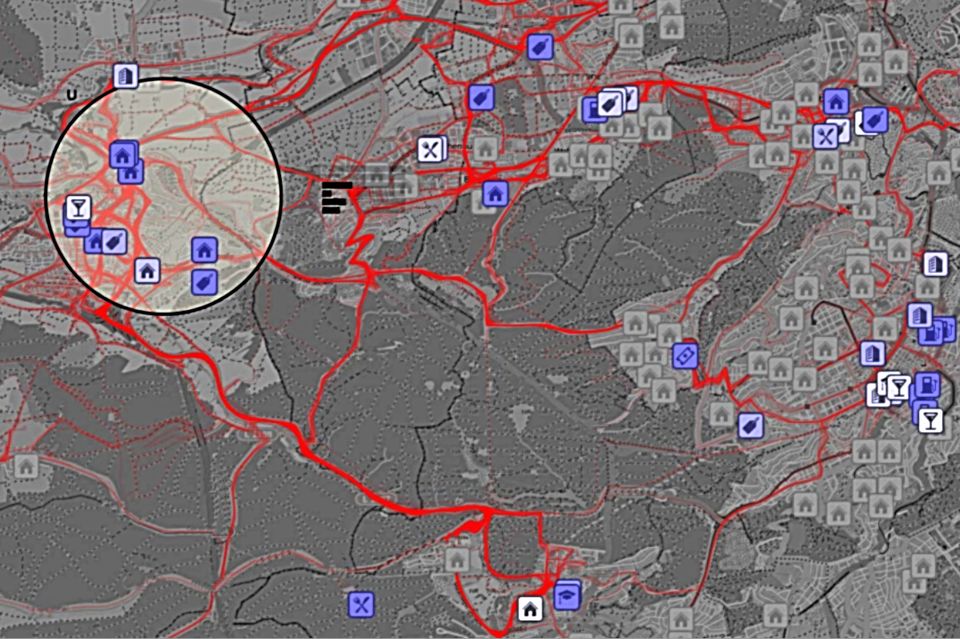}\label{fig:GISlumi}}\\
	\end{tabular}
	\caption{Examples of color images (left column) improved by sharpening (right column) with virtual viewing distances of 49\,cm for 
  vortex visualization with scattering (row 1), 48\,cm for volume shadowing (row 2), and 32\,cm for GIS data (row 3). Images in the left column (top to bottom): \textcopyright\space~IEEE. Reprinted, with permission, from Ament et al.~\shortcite{Ament:VIS:2013,Ament:VIS:2014} and Krueger et al.~\shortcite{Krueger:pvis:2014}.}
	\label{tbl:colorImages}
	\vspace{-1em}
\end{figure*}

\rev{Figures~\ref{fig:tomato} and~\ref{fig:sliceResults} are examples of grayscale volume slices with perceptual colormaps applied. The tomato and the hurricane datasets are generated using  ColorBrewer color maps~\cite{Harrower:CJ:2003}, while the MRI scan is encoded with a multihue isoluminant color map~\cite{Kindlmann:2002:FLM:602099.602145}.}
Examples of RGB-colored images from previously published visualization techniques are shown in Figures~\ref{fig:introImg} and~\ref{tbl:colorImages}.

The tomato CT scan (Figure~\ref{fig:tomato}), which has been discussed in Section~\ref{sec:overcomp}, and the MRI brain scan (Figure~\ref{fig:sliceResults}, row 1) contain rather clear boundaries between anatomically meaningful structures.
Without sharpening, the MRI image has a washed-away look so that one cannot easily separate the brain from the surrounding tissues (cyan) and it is difficult to recognize the delicate folded details. 
Fine details of the brain become clearly noticeable with a viewing distance of 75\,cm.
Figure~\ref{fig:sliceResults}, row 2, shows the pressure attribute of the Hurricane Isabel dataset of one time step that contains smooth and homogeneous structures of large scale with subtle yet important vortex details. 
The hurricane eye, the spiral arms, and the shore area are roughly visible in the original visualization; with a viewing distance of 90\,cm, the eyewall feature and spiral structures are more visible.

Depth-of-field volume renderings~\cite{Mathias:cgf11} of a combustion simulation are shown in Figure~\ref{fig:introImg}.
Figure~\ref{tbl:colorImages}, row 1, shows volume rendered images generated by a volumetric scattering method \cite{Ament:VIS:2013}. 
The original image contains sharp edges but loses some fine details inside vortices.
In contrast, using our method with a viewing distance of 49\,cm, fine details inside vortices are enhanced, and the image becomes crisper.
\rev{A lowpass volumetric shadowing technique~\cite{Ament:VIS:2014} is able to enhance the depth cues of a volume rendering as shown in Figure~\ref{tbl:colorImages}, row 2.
	The edges in the original rendering are fuzzy. 
	Enhanced with luminance at a viewing distance of 48\,cm, object boundaries and high-frequency details are highlighted and become clearly visible.}
The third row of Figure~\ref{tbl:colorImages} shows a focus-and-context visual analysis method for movement behavior~\cite{Krueger:pvis:2014} applied to a GIS dataset of a city: dark red road networks and the region inside the circle are in focus, while other regions have reduced contrast.
The sharpened result generated with a viewing distance of 32\,cm enhances the overall contrast while preserving the focus-and-context impression.
Inside the circle of focus, icons and structures become more prominent and are emphasized by halos; outside the focus, one could gain insights more easily with slightly enhanced details that are not distracting users from the focus region.

\section{Conclusion and Future Work}
\label{sec:conclusion}

We propose an image-based method that compensates contrast loss depending on viewing distance.
We start from a well-accepted computational perception model~\cite{PattanaikFFG98}.
Then, we simplify the model with a spectral approximation to invert the contrast loss due to viewing distance by compensating the power spectrum.
Specifically, we extract bandpass images and find the optimal weighting for these images that compensate the power spectrum by solving an optimization problem.
Compensation and overcompensation can be easily achieved with a simple user interface.
A wide range of datasets that have representative image features is used as examples to demonstrate the usefulness of our method.

\rev{Our method has some limitations. 
In particular, the white noise approximates behaviors of the mean of visualization images, but might deviate from individual input datasets. Although all 50 images for spectral analysis are carefully hand-picked to represent typical 
visualizations, it is still a rather small number and may not represent all variations. 
}

For future work, we would like to extend the method for automatic visualization sharpening in VR/AR environments by setting the viewing distance with sensors for an immersive experience.
\rev{
In addition, a larger number of visualization images could be used for spectral analysis for better calibration.
Finally, a more accurate model that inverts the perceptual simulation~\cite{PattanaikFFG98} could be devised to convey more faithful visualizations to users.}

\begin{acks}
This work is funded by the Deutsche Forschungsgemeinschaft (DFG, German Research Foundation) -- Project-ID 251654672 -- TRR 161, National Institute of General Medical Sciences of the National Institutes of Health under grant number P41 GM103545-18, and the Intel Graphics and Visualization Institutes.
\end{acks}

\bibliographystyle{ACM-Reference-Format}
\bibliography{contrastEnhModel}
\end{document}


\title{Spectral Visualization Sharpening---Supplemental Material}

\author{Liang Zhou}
\affiliation{%
  \department{SCI Institute}
  \institution{University of Utah}}

\author{Rudolf Netzel}
\affiliation{%
	\department{VISUS}
	\institution{University of Stuttgart}}

\author{Daniel Weiskopf}
\affiliation{%
	\department{VISUS}
  \institution{University of Stuttgart}}

\author{Chris R. Johnson}
\affiliation{%
  \department{SCI Institute}
  \institution{University of Utah}}

\renewcommand{\shortauthors}{Zhou, Netzel, Weiskopf, and Johnson}

\begin{abstract}
	This supplemental material contains contents left out from the paper for conciseness. Specifically, we provide sources of visualization images used in the spectrum study, the actual simulated sequence using the perception model and the power spectra of one visualization. We further show the derivation of the spectral model (Equations 1 and 2) in Section 3.3 of the paper. 
	

\end{abstract}  
\maketitle
\section{Visualizations Used In Spectrum Analysis}
We obtained 50 visualizations from the Internet using the Google image search engine and manual selection to cover a wide range of visualization examples. These images can be classified into five categories: volume rendering, flow visualization, DTI visualization, GIS visualization, and slices of medical scans (CT and MRI).

\begin{figure}[htb]
	\subfloat[Original]{\includegraphics[width=0.49\linewidth]{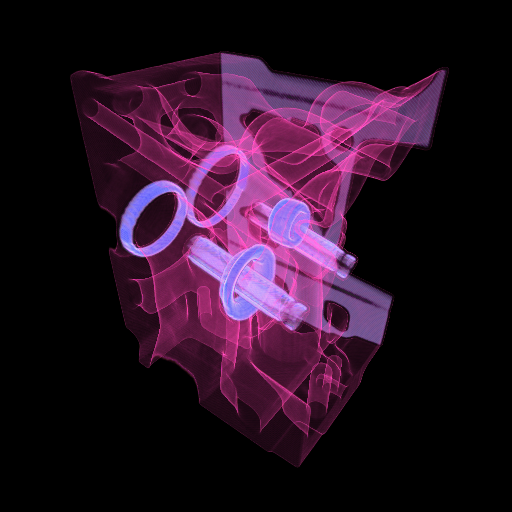}}\hfill
	\subfloat[Sharpened]{\includegraphics[width=0.49\linewidth]{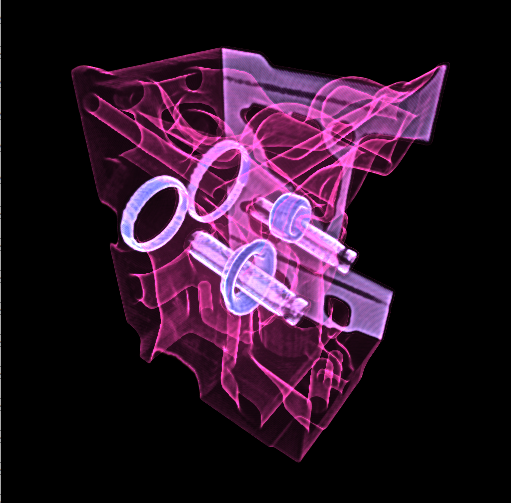}}
	\caption{The original rendering (a) and the sharpened visualization (b) of the Engine dataset. Image (a) \textcopyright\space S. Roettger. Reprinted, with permission, from \url{http://schorsch.efi.fh-nuernberg.de/roettger/uploads/Main/Engine.png}.}
	\label{fig:engineSharp}
\end{figure}
\subsection{Data Sources}
We hereby list sources of these visualization images.
\subsection*{Volume Rendering}

\sloppy
\begin{itemize}
\item\url{https://www.marcusbannerman.co.uk/articles/img/Bonsai.jpg}
\item\url{http://schorsch.efi.fh-nuernberg.de/roettger/uploads/Main/Engine.png}
\item\url{http://old.cescg.org/CESCG-2004/web/Bruckner-Stefan/html/}
\item\url{https://www.exxim-cc.com/img/Shoulder_vr.jpg}
\item\url{https://www.exxim-cc.com/img/Volume_rendered_foot_p.jpg}
\item\url{https://www.gcc.tu-darmstadt.de/media/gcc/projects/volcomp/xmas_orig_light_529x0.jpg}
\item\url{https://www.tue.nl/en/research/research-groups/medical-image-analysis/people/BRomeny/PACS/NERI_09-ter Haar Romeny_files/image013.jpg}
\item\cite{Kruger:vis06}
\item\cite{PRH09}
\end{itemize}

\subsection*{Flow Visualization}
\begin{itemize}
\item\url{https://www3.nd.edu/~cwang11/2dflowvis.html}
\item Slides 21, 41 of\\ \url{https://www3.nd.edu/~cwang11/research/vis13-tutorial-weiskopf.pdf}
\item\cite{Sadlo2014LyapunovTime}
\item\cite{Lawonn:Eurovis14}
\item\cite{Treib:tvcg12}
\item\cite{Karch2012Dye}
\end{itemize}

\subsection*{GIS Visualization}
\begin{itemize}
\item\url{http://vis.pku.edu.cn/trajectoryvis/en/densitymap.html}
\item\url{https://carto.com/blog/location-intelligence-end-of-gis-as-we-know-it}
\item\url{http://www.cs.rug.nl/svcg/uploads/Shapes/SBEB_frair.jpg}
\item\cite{Scheepens:PVis11}
\item\cite{Wu:VIS16}
\item\cite{Wang:TVCG2014}
\item Slides 6, 29, 39, 43 of 
\url{http://vis.cs.kent.edu/TrajAnalytics/files/Overview of Visualization.pdf}
\end{itemize}

\begin{figure*}[htb]
	\subfloat[$d=10$]{\includegraphics[height = \tomatoHeight]{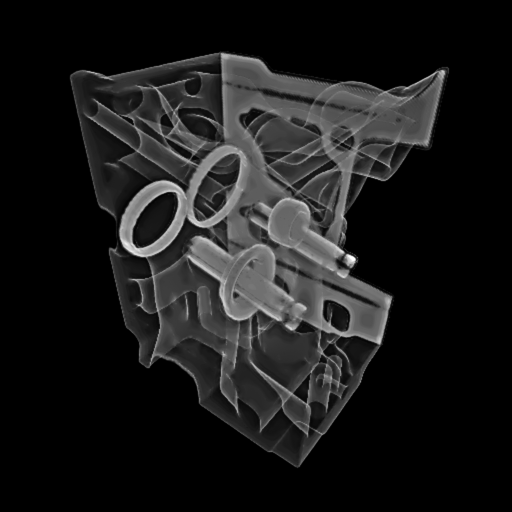}}
	\subfloat[$d=20$]{\includegraphics[height = \tomatoHeight]{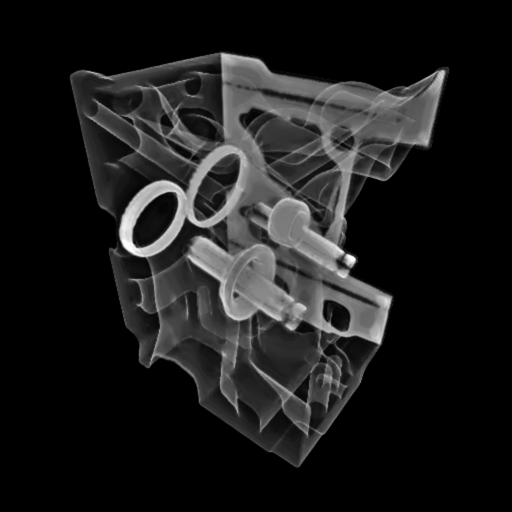}}
	\subfloat[$d=30$]{\includegraphics[height = \tomatoHeight]{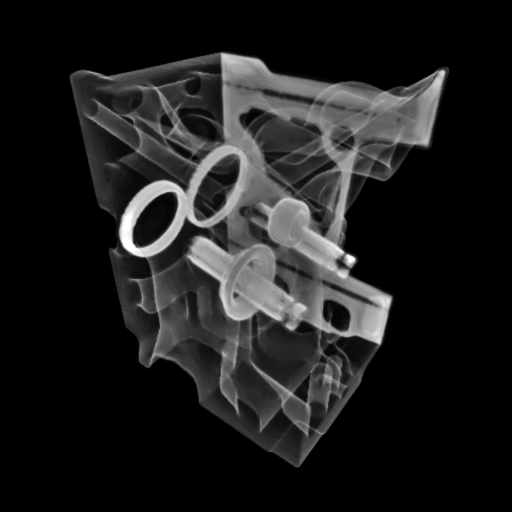}}
	\subfloat[$d=40$]{\includegraphics[height = \tomatoHeight]{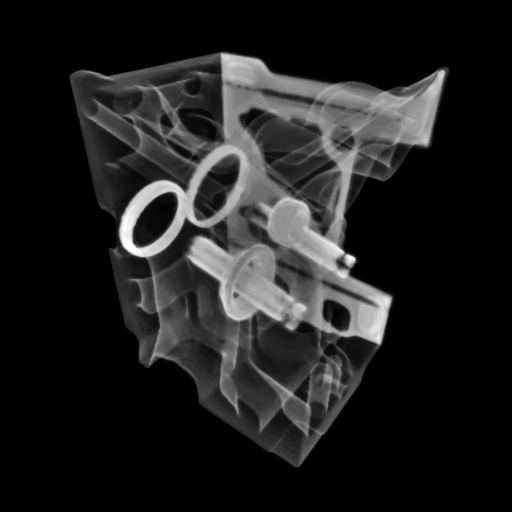}}
	\subfloat[$d=50$]{\includegraphics[height = \tomatoHeight]{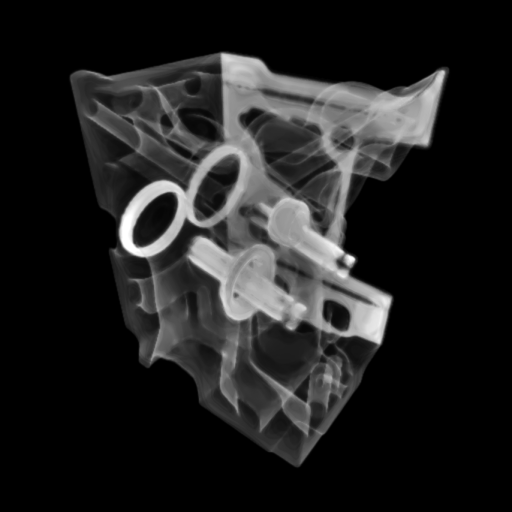}}\\
	\subfloat[$d=60$]{\includegraphics[height = \tomatoHeight]{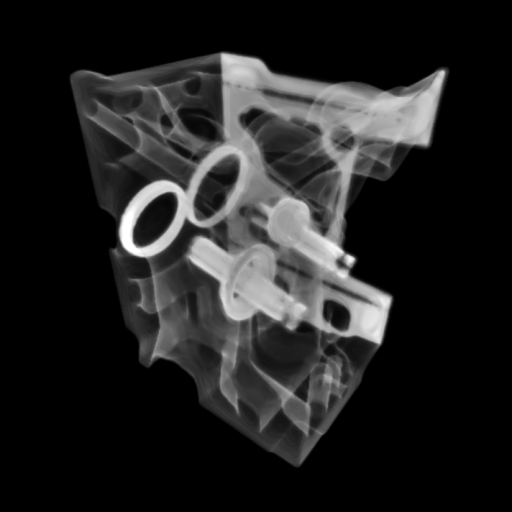}}
	\subfloat[$d=70$]{\includegraphics[height = \tomatoHeight]{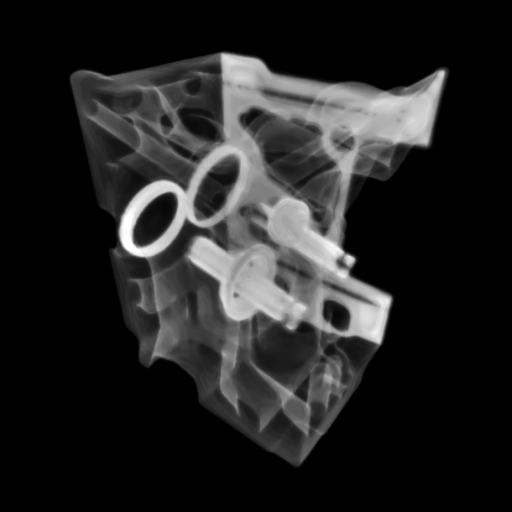}}
	\subfloat[$d=80$]{\includegraphics[height = \tomatoHeight]{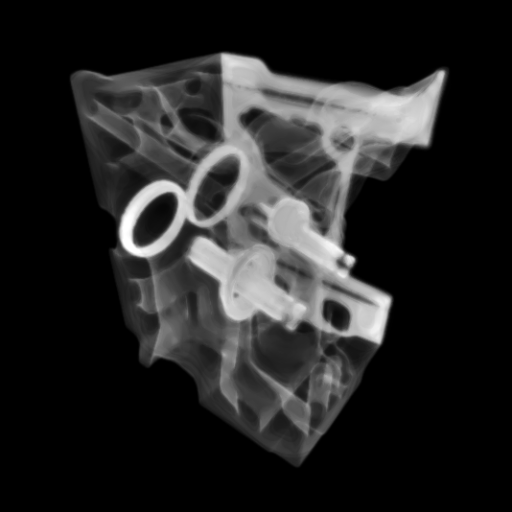}}
	\subfloat[$d=90$]{\includegraphics[height = \tomatoHeight]{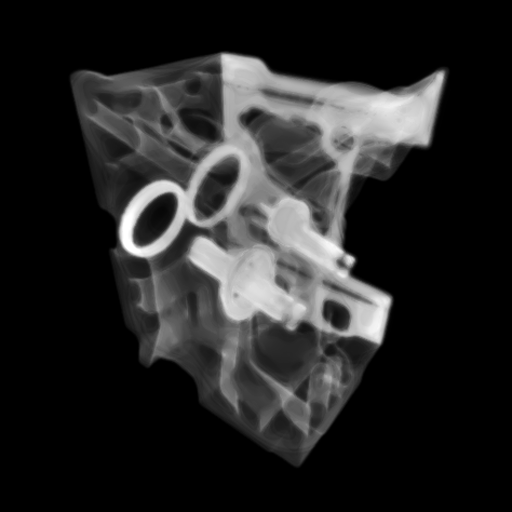}}
	\subfloat[$d=100$]{\includegraphics[height = \tomatoHeight]{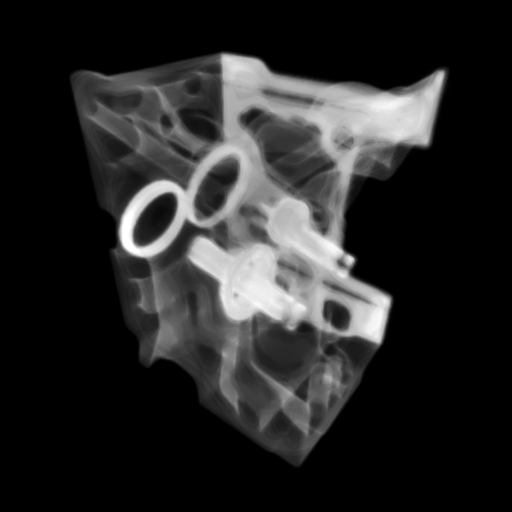}}
	\caption{Simulations of the Engine data with viewing distance $d=10$\,cm to $d=100$\,cm.}
	\label{fig:simSeq}
\end{figure*}
\subsection*{DTI Visualization}
\begin{itemize}
\item\url{https://www.siemens-healthineers.com/nl/magnetic-resonance-imaging/options-and-upgrades/clinical-applications/syngo-dti-tractography}
\item\url{https://www.researchgate.net/profile/Donna_Sorkin2/publication/308203750/figure/fig4/AS:667817812701207@1536231582517/DTI-image-showing-central-brain-tracts_W640.jpg}
\item\url{https://s3.amazonaws.com/spectrumnews-web-assets/uploads/image-archive/images/images-2014/blog2014/20141223blogdtierrors.jpg}
\item\url{https://qph.fs.quoracdn.net/main-qimg-e24e3fe93bd87b64c2a0ae9e0571b126.webp}
\item\cite{FRIEDMAN20141039}
\item\cite{Zhukov:Vis03}
\item Figures 4, 10, 12, 16 of~\cite{Isenberg:2015}
\end{itemize}

\subsection*{Medical Imaging}
\begin{itemize}
\item\url{http://zoi.utia.cas.cz/files/u3/ct.png}
\item\url{https://d2htw67fsvc5r2.cloudfront.net/blog/wp-content/uploads/2018/04/brainMRI.png}
\item\url{https://www.ctisus.com/resources/library/teaching-files/ctpet/391724.jpg}
\item\url{https://d2htw67fsvc5r2.cloudfront.net/blog/wp-content/uploads/2018/04/MRIabdomen.jpg}
\item\url{https://prod-images.static.radiopaedia.org/images/32248663/5931fb2a0d5f599957ba71ff5e5707_big_gallery.jpeg}
\item\url{http://www.radiologyassistant.nl/data/bin/a5097978bbd183_6b.jpg}
\item\url{https://d2htw67fsvc5r2.cloudfront.net/blog/wp-content/uploads/2018/04/brainMRI.png}
\item\url{https://www.materprivate.ie/__uuid/150fac2e-54f5-4d70-866f-e348f8800a88/MRI-Head2.jpg}
\item\url{https://my-ms.org/images/CerebCor 201.16.jpg}
\item\url{http://www.brainfacts.org/-/media/Brainfacts2/In-the-Lab/Tools-and-Techniques/Article-Images/}
\end{itemize}

\subsection{Analysis of One Visualization}

In Figure~\ref{fig:simSeq}, we show an actual sequence of simulations with viewing distance from 10\,cm to 100\,cm of the volume rendering of the engine data shown in Figure~\ref{fig:engineSharp}(a). 

Applying our visualization sharpening to the Engine data with a viewing distance of 80\,cm yields the result in Figure~\ref{fig:engineSharp}(b), and the original image is shown in Figure~\ref{fig:engineSharp}(a) for comparison.

\begin{figure}[htb]
	\subfloat[]{\includegraphics[width= 0.9\linewidth]{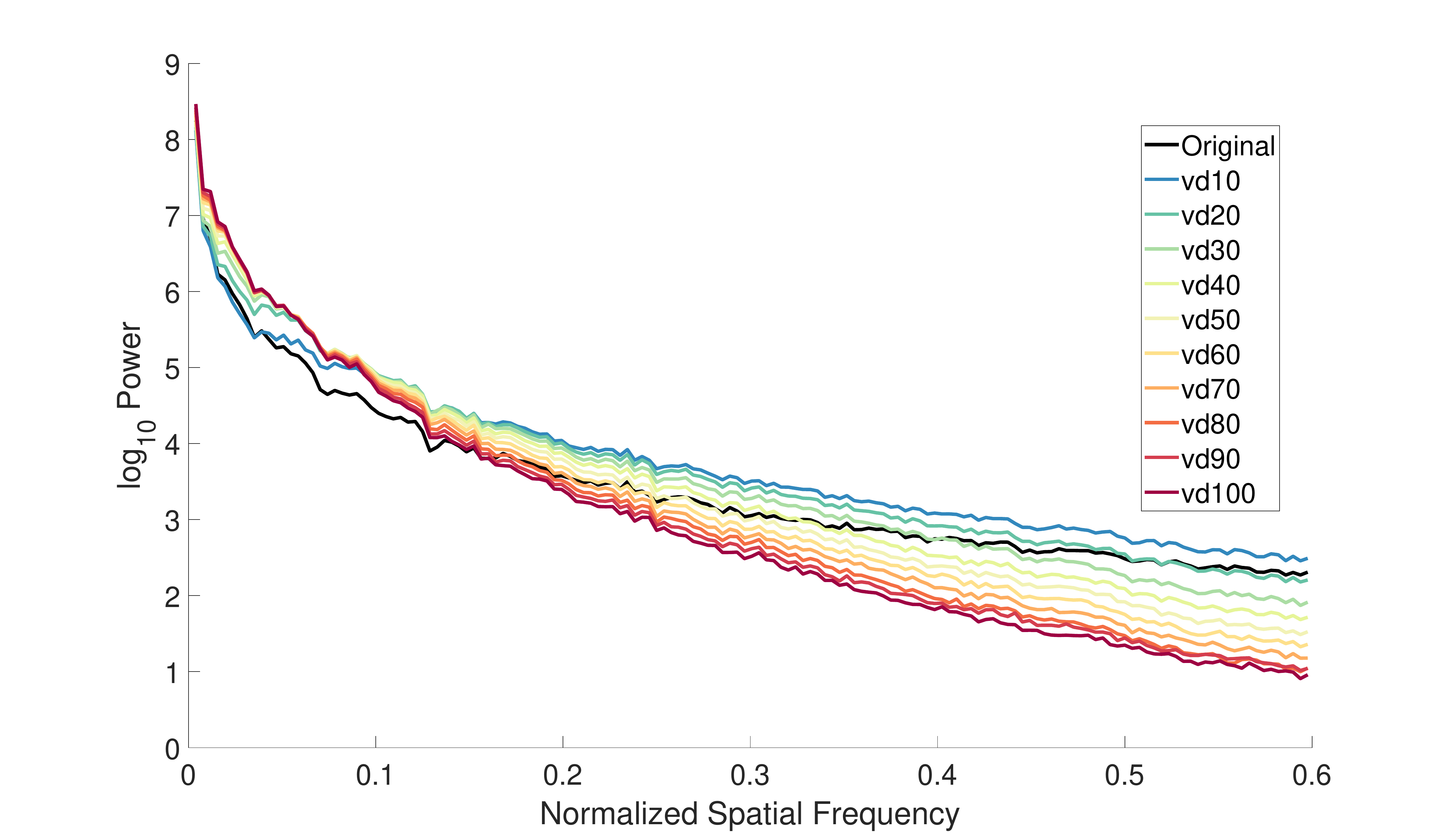}}\\
	\subfloat[]{\includegraphics[width= 0.9\linewidth]{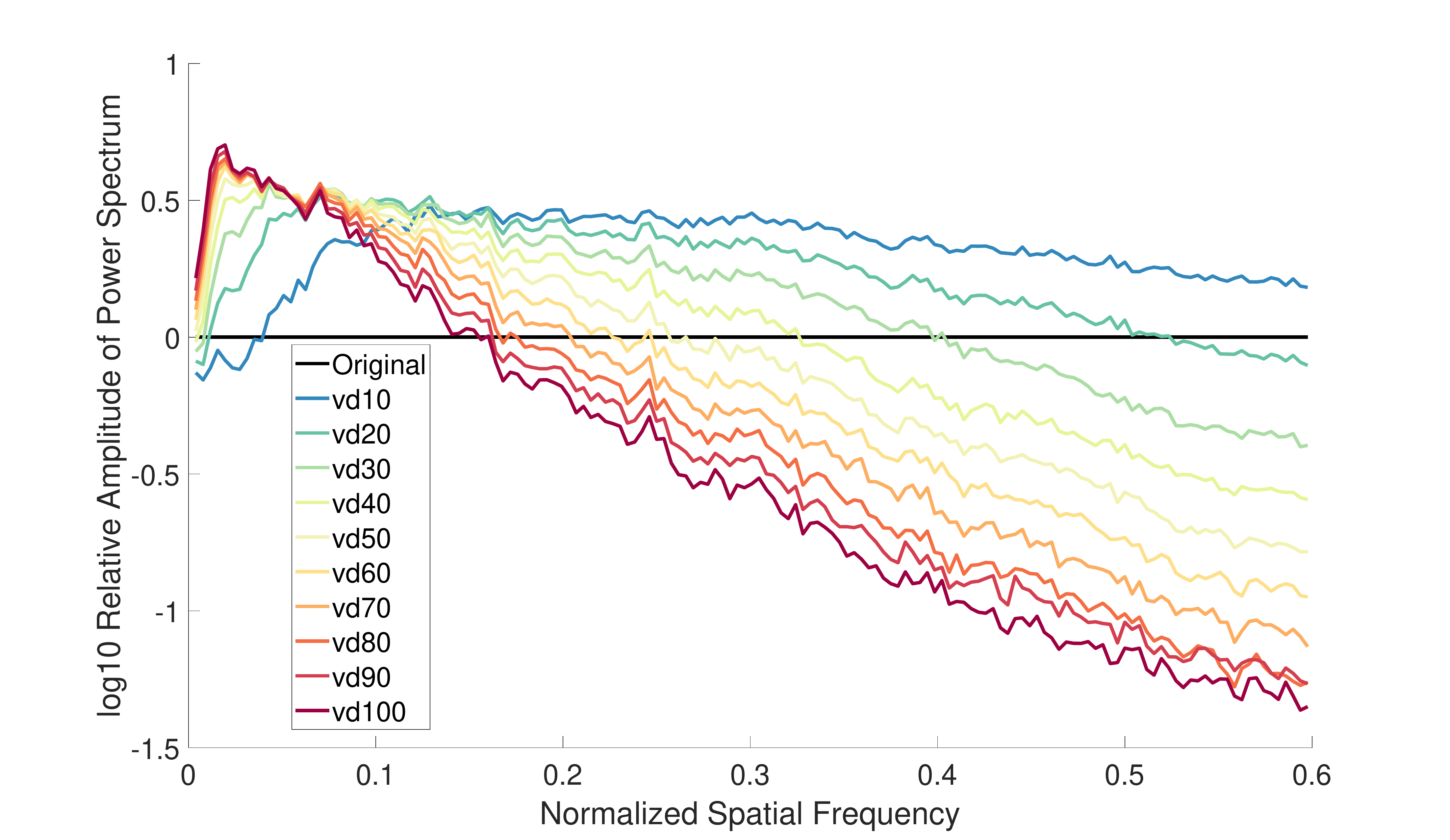}}
	\caption{Logarithmic power spectra of perception simulations of the Engine data (a), and (b) logarithmic relative amplitude of power spectra of simulations normalized by the power spectrum of the original image. }
	\label{fig:engineSpec}
\end{figure}

The logarithmic power spectra of the Engine dataset and the logarithmic relative amplitude curves of these power spectra can be seen in Figure~\ref{fig:engineSpec}.
Note that the similarity between Figure~\ref{fig:engineSpec}(b) and the logarithmic relative amplitude of white noise in Figure 4 in the paper: the curves have similar behaviors over the frequency range and also have similar logarithmic relative amplitude.

\section{Derivation of the Spectral Model (Equations 1 and 2 in the Paper)}
We have a filtering that is frequency-dependent, therefore, we formulate a spectral perception model for an image $f(x,y)$:
\begin{equation}
S_d[f(x,y)] = \mathscr{F}^{-1}[\mathscr{F}[f(x,y)] \cdot H_{d}(\nu)] \;,
\label{eqn:simpModel}
\end{equation}
where $S_d[\cdot]$ is the perceptual simulation operation for a virtual viewing distance $d$, and $H_d$ is a transfer function. 
As shown in Equation~\ref{eqn:transfunc}, $H_d$ is the power spectrum of the white noise data simulated at $d$ (one of the curves shown in Figure 4(b) in the paper):
\begin{align}
	H_d(\nu) = P_r[\mathscr{F}[S_d[n(x,y)]]](\nu)\;.
	\label{eqn:transfunc}
\end{align}
Here, we assume that the original noise image $n(x,y)$ is normalized in the sense that its power spectrum averages to one.

We can replace the input image by the sum of a number of bandpass images; therefore, Equation~\ref{eqn:simpModel} can be rewritten as:
\begin{align}
	S_d[f(x,y)] &= \mathscr{F}^{-1}\left[\sum_{i=1}^{L}\mathscr{F}[f_i(x,y)] \cdot H_{d}(\nu)\right]\;.
\end{align}
Since the transfer function changes in a controlled way, it is valid to approximate the transfer function on each frequency interval of the bandpass images with a constant $H_i$.
The Fourier transform is a linear operator, therefore we have:
\begin{align}
	\centering
	S_d[f(x,y)] &\approx \mathscr{F}^{-1}\left[\sum_{i=1}^{L}\mathscr{F}[f_i(x,y)] \cdot H_i\right] \\
	&= \sum_{i=1}^{L}H_i\cdot f_i(x,y)\;.\label{eqn:simpModelApprox}
\end{align}
This equation serves as the mathematical model of our simplified perception pipeline.

Then, it is possible to inverse the model for a virtual viewing distance $d$, by replacing the constants $H_i$ with their inverse $1/H_i$ in Equation~\ref{eqn:simpModelApprox}. 
Effectively, it raises the power spectrum of the perceived image at $d$ to the constant value leading to perceptual compensation.

\bibliographystyle{ACM-Reference-Format}
\bibliography{contrastEnhModel}